\documentclass[aps,prb,reprint]{revtex4-1}
\usepackage{graphicx,color,bm,amsmath,amssymb,url,natbib}

\begin{document}

\title{Activated vortex lattice transition in a superconductor with combined sixfold and twelvefold anisotropic interactions}

\author{D.~Minogue}
\affiliation{Department of Physics and Astronomy, University of Notre Dame, Notre Dame, Indiana 46656, USA}

\author{M.~R.~Eskildsen}
\altaffiliation{email: eskildsen@nd.edu}
\affiliation{Department of Physics and Astronomy, University of Notre Dame, Notre Dame, Indiana 46656, USA}

\author{C.~Reichhardt}
\author{C.~J.~O.~Reichhardt}
\affiliation{Theoretical Division, Los Alamos National Laboratory, Los Alamos, New Mexico 87545, USA}

\date{\today}

\begin{abstract}
Numerical simulations are used to examine the transition dynamics between metastable and ground state vortex lattice phases in a system with combined sixfold and twelvefold contributions to the vortex-vortex interactions.
The system is initially annealed using a twelvefold anisotropy, yielding domains of two different orientations and separated by grain boundaries. 
The vortex-vortex interaction is then suddenly changed to a sixfold anisotropy, rendering the twelvefold state metastable.
Applying a drive that mimics an oscillating magnetic field causes the metastable state to decay, indicated by the structure factor that evolves from twelve to six peaks.
The results fit the behavior seen in recent small-angle neutron scattering studies of the vortex lattice in MgB$_2$.
At higher drive amplitudes, the decay exhibits a two step process in which the initial fast decrease is followed by a slower regime where avalanches or bursts are correlated with dislocation annihilation events. 
The results are compared to other types of metastable systems with quenched disorder that decay under a periodic external drive.
\end{abstract}

\maketitle

\section{Introduction}
Magnetic flux enters a type-II superconductor in the form of quantized vortices~\cite{Abrikosov:1957vu}.
In the absence of strong pinning by impurities or thermally driven disordering, the mutual repulsion between vortices causes them to crystallize into an ordered vortex lattice (VL)~\cite{Blatter:1994gz}.
In a perfectly isotropic superconductor the VL would be triangular and without a preferred orientation relative to the host lattice \cite{Kleiner:1964ih};
however, real materials always posses some degree of anisotropy that causes an alignment of the VL relative to the crystalline axes, often giving rise to field and/or temperature driven symmetry transitions~\cite{Muhlbauer:2019jt}.
Such transitions lead to the formation of competing VL domains with different orientations. 
The VL can thus serve as a simple model system for domain nucleation and growth, with relevance for a wide range of physical systems including martensitic phase transitions~\cite{Wang:2017jw}, domain switching in ferroelectrics~\cite{Shin:2007gu}, and the skyrmion lattice in chiral magnets~\cite{Makino:2017hh,Nakajima:2017uc,Bannenberg:2017ws}.

For magnetic fields applied along the $c$ axis of superconductors with a hexagonal crystal structure, the VL is empirically found always to have a triangular symmetry, but will often undergo continuous and sometimes even non-monotonic rotations as seen in MgB$_2$~\cite{Cubitt:2003aa} and UPt$_3$~\cite{Huxley:2000aa,Avers:2020wx}.
In addition, the VL in these materials exhibit significant metastability when cooled or heated across equilibrium phase transitions in a constant field~\cite{Huxley:2000aa,Das:2012cf}.
To reach the equilibrium configuration the VL must be excited by an external perturbation, which causes vortex motion and allows the system to find the global minimum in the free energy~\cite{Muhlbauer:2019jt}.
In MgB$_2$ the transition from the metastable to the equilibrium state has been studied extensively using small-angle neutron scattering (SANS), revealing an activation barrier that increases as the metastable state is gradually suppressed~\cite{Louden:2019wx}.
This ``work hardening'' is speculated to arise from a proliferation of grain  boundaries in the VL.
Although spatially resolved SANS measurements provide some support for such a scenario~\cite{Louden:2019jn}, a real space study of the VL is desirable.

Spatial information about the VL can be obtained by molecular dynamics (MD) methods in which the vortices are treated as point particles with repulsive interactions.
In addition to static VL configurations, MD simulations provide information about the dynamics of a large number of vortices over long times.
This technique has been widely used to study how pinning affects the structural~\cite{Shi91} and dynamical~\cite{Olson98a,Fangohr01,Klongcheongsan:2009fi} properties of vortex matter.
Due, however, to the added complexity introduced by a fully anisotropic interaction potential and the resulting nonradial forces, such studies were limited to systems with isotropic vortex-vortex interactions.
Recently we implemented a phenomenological model that allows MD simulations of the VL in the presence of anisotropic interactions~\cite{Olszewski:2018fp,Olszewski:2020jy,Roe:2022tl}.
This was used to study systems with combined sixfold and twelvefold anisotropies relevant for MgB$_2$~\cite{Olszewski:2020jy}.
By varying the ratio of the six- and twelvefold contributions to the interaction potential, it was possible to reproduce the three different triangular VL phases observed experimentally in this material~\cite{Cubitt:2003aa,Das:2012cf}.
The VL phases, denoted F, L and I, differ only in their orientation relative to the MgB$_2$ crystalline axes.
Specifically, F and I phases are oriented along the two high symmetry directions within the hexagonal basal plane ($\bm{a}$ along the nearest neighbor direction and $\bm{a}^*$ along the next nearest neighbor direction), and are connected by an intermediate, continuously rotating L phase~\cite{Cubitt:2003aa,Hirano:2013jx}.
The orientation of the VL phases relative to $\bm{a}$ and $\bm{a}^*$ is shown in Ref.~\onlinecite{Hirano:2013jx}, and the directions of the crystalline axes relative to the interaction potential are indicated in Fig.~\ref{fig1}(a).
The simulations also spontaneously produced VL dislocations as well as a range of vacancy and interstitial defects that were analyzed in terms of their energy cost~\cite{Olszewski:2020jy}.

In this work we extend our MD simulations to study the gradual VL transition from a metastable L phase towards the equilibrium F phase.
The metastable L phase is obtained by first annealing the system with a vortex-vortex interaction potential that has a twelvefold anisotropy and subsequently changing this to a sixfold anisotropy. 
Weak pinning is included to stabilize the initial twelvefold symmetric phase, and the temperature is kept low enough that thermal fluctuations alone are insufficient to overcome the activation barrier for a transition to the ground state.
We do not observe metastable effects in simulations that does not include pinning, suggesting  that at least some weak quenched disorder plays a role in the SANS experiments~\cite{Das:2012cf,Louden:2019wx,Louden:2019jn}.
Here, the quenched disorder will decrease thermal hopping effects so that only the mechanical driving from changing the effective field, as discussed below, will play a role. 

In the SANS experiments, the transition to the ground state is driven by small-amplitude magnetic field oscillations, leading to a temporary variation in the vortex density akin to a breathing mode~\cite{Louden:2019wx}.
Such a scenario is not straightforward to implement in the MD simulations, due to the periodic boundary conditions, and we instead employ a periodic variation of the superconducting penetration depth.
This changes the effective vortex-vortex interaction strength and thus mimics a field oscillation.
As successive penetration depth cycles are applied, the metastable L phase gradually rotates towards the ground state F phase.
Initially, the angle between the two degenerate L phase domain orientations decreases as a power law with an exponent that increases with the amplitude of the penetration depth cycles, similar to what was observed in the SANS experiments~\cite{Louden:2019wx}.
For the larger amplitudes there is a cross-over to a logarithmic behavior as the number of penetration depth cycles increases.
Here, the VL relaxation occurs in bursts or avalanches that are associated with correlated dislocation annihilation events along the grain boundaries.
We discuss how these behaviors compare to two-dimensional (2D) coarsening models~\cite{Elder92,Purvis01,Boyer02,Boyer04}, and show that the $\log(t)$ decay could arise due to coarsening or glassy motion with quenched disorder~\cite{Reichhardt06a}.

Our approach can be applied to other particle-based systems with anisotropic interactions, including skyrmion lattices~\cite{Jonietz:2010jz,White:2014ji,Zhang:2018bg} or colloidal particles with anisotropic interactions~\cite{Eisenmann:2004iv,Glotzer:2007uo}.
Furthermore, it will allow modeling of the dynamics associated with the generation and recombination of dislocations, grain boundaries and domain formation both in the VL~\cite{Louden:2019wx,Louden:2019jn} and in general~\cite{Irvine:2013dk,Lavergne:2018ds}.
We discuss the relevance of our results to the decay of metastable states in other systems with periodic perturbations, such as granular matter~\cite{Nowak98,Gago20}, colloidal systems~\cite{Reichhardt06a}, magnetic systems with return point memory~\cite{Pierce05,Libal12},
and other types of periodically sheared systems that relax to a steady state under ac driving~\cite{Valenzuela02,Corte08,Pasquini21}.

\section{Methods}
\label{SecMethods}
\begin{figure}[b]
    \includegraphics{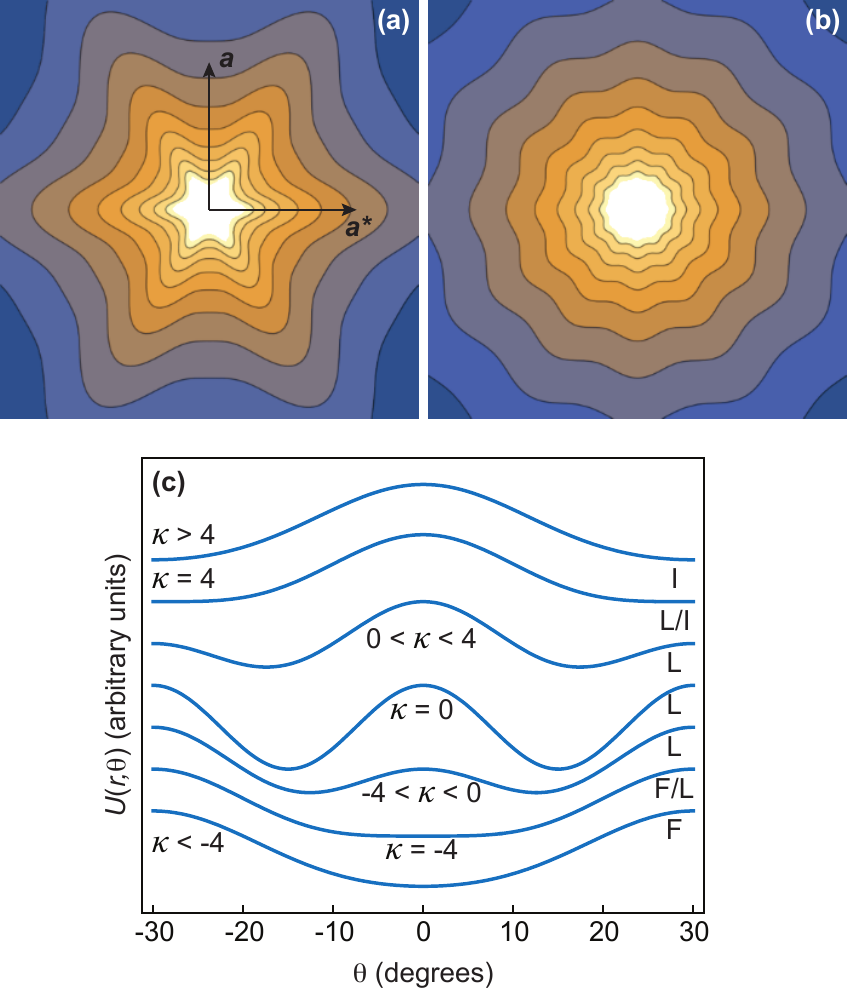}
    \caption{\label{fig1}
        Vortex-vortex interaction potential given by Eq.~(\ref{K612potential}).
        Heat maps illustrate the spatial variation of $U(r,\theta)$ for $\kappa = 5$ (a) and $\kappa = 0$ (b).
        In both cases $K_{12}=0.0125$.
        Arrows in (a) indicate the directions of the MgB$_2$ in-plane crystalline axes relative to the interaction potential.
        (c) Angular dependence of $U(r,\theta)$ for a fixed radius and different values of $\kappa$.
        For each curve, the corresponding ground state VL phase is indicated on the right.
        The FL- and LI-transitions occur at $\kappa = \pm 4$.}
\end{figure}

We follow the approach of Olszewski {\em et al.} and use a phenomenological vortex-vortex interaction potential that includes six- and twelvefold anisotropies~\cite{Olszewski:2020jy}:
\begin{equation}
    U(r,\theta) = A_v K_0(r) \left[ 1 + K_6 \cos^2 3 \theta + K_{12} \cos^2 6 \theta \right].
    \label{K612potential}
\end{equation}
Here $r$ and $\theta$ are the distance and angle, respectively, between two vortices at positions $\bm{r}_i$ and $\bm{r}_j$, namely $r = |\bm{r}_i - \bm{r}_j|$ and $\theta = \tan^{-1}(r_y / r_x)$, with $\bm{r} = \bm{r}_i - \bm{r}_j$, $r_x = \bm{r} \cdot \bm{\hat{x}}$, and $r_y = \bm{r} \cdot \bm{\hat{y}}$.
The radial part of the repulsive interaction potential is given by the amplitude $A_v$ times the zeroth order Bessel function $K_0(r)$~\cite{Tinkham:1996un}.
The parameters $K_6$ and $K_{12}$ are the magnitudes of the six- and twelvefold contributions to the potential, with an anisotropy ratio defined by $\kappa = K_6/K_{12}$.
When $|\kappa| > 4$, the potential is dominated by the six-fold term, with minima, for a fixed $r$, along either the horizontal ($\kappa < -4$) or vertical ($\kappa > 4$) direction and at every $60^{\circ}$.
This is illustrated in Fig.~\ref{fig1}(a).
For $|\kappa| \leq 4$, there are 12 minima in the interaction potential as shown in Fig.~\ref{fig1}(b), and consequently there are two degenerate triangular VL configurations.
The locations of the minima change as $\kappa$ is varied and are given by
\begin{equation}
  \theta_{\text{min}}(\kappa) = \pm \frac{1}{6} \arccos \left( - \frac{\kappa}{4} \right).
  \label{thetamin}
\end{equation}
Six of the minima rotate clockwise as $\kappa$ is reduced while the other six rotate counterclockwise, as shown in Fig.~\ref{fig1}(c).
Changing the sign of $\kappa$ is equivalent to a $30^{\circ}$ rotation, which exchanges the locations of the minima and maxima.

It should be noted that the interaction in Eq.~(\ref{K612potential}) is not based on an {\em ab-initio} derivation, and, within Ginzburg-Landau (GL) theory, it is strictly speaking not possible to formulate an interaction potential for a single vortex~\cite{Silaev.2018}.
Instead Eq.~(\ref{K612potential}) is motivated by an expression for the entire VL free energy obtained from GL theory~\cite{Zhitomirsky:2004aa}.
As shown by Olszewski {\em et al.}, this potential reproduces the experimentally observed macroscopic VL phases in MgB$_2$~\cite{Cubitt:2003aa} and UPt$_3$\cite{Huxley:2000aa,Avers:2020wx} for a suitable choice of parameters $K_6$ and $K_{12}$~\cite{Olszewski:2020jy}.
It can thus be considered a minimal and conceptually simple phenomenological model for MD simulations of the VL in superconductors with a hexagonal crystal structure.

In the MD simulations the dynamics of each vortex is governed by an overdamped equation of motion:
\begin{equation}
  \eta \frac{d\bm{r}_i}{dt} = \bm{F}_i^{vv} + \bm{F}_i^p +\bm{F}_i^T.
\end{equation}
Here $\eta$ is the damping constant which is set equal to unity.
The force field from the surrounding vortices is given by
$\bm{F}^{vv} = -\bm{\nabla}(U) = (-\partial U/\partial x, -\partial U/\partial y)$, yielding components
\begin{widetext}
  \begin{subequations}
    \begin{eqnarray}
      \frac{F_x}{A_v}
        & = & \cos \theta \, K_1(r) \left[ 1 + K_6 \cos^2 \theta + K_{12} \cos^2 6 \theta \right] 
              - \frac{1}{r} \sin \theta \, K_0(r) \left[ 3 K_6 \sin 6 \theta + 6 K_{12} \sin 12 \theta \right] \label{Fvv_a}\\
      \frac{F_y}{A_v}
        & = & \sin \theta \, K_1(r) \left[ 1 + K_6 \cos^2 3 \theta + K_{12} \cos^2 6 \theta \right]
              + \frac{1}{r} \cos \theta \, K_0(r) \left[ 3 K_6 \sin 6 \theta + 6 K_{12}\sin 12 \theta \right] \label{Fvv_b}.
    \end{eqnarray}
  \end{subequations}
\end{widetext}
We consider a system of size $L \times L$ with $L = 108\lambda$, and apply periodic boundary conditions in both the $x$ and $y$ directions.
Here $\lambda$ is the London penetration depth.
The sample contains $N_v = 5130$ vortices interacting with $N_p = 4000$ pinning sites placed randomly in non-overlapping positions and generating a pinning force $\bm{F}_i^p$.
Each pinning site is modeled as a parabolic trap of range $r_p = 0.25$ that can exert a maximum pinning force of $F_p = 1.15$ on a vortex.
The vortex-pin interaction is isotropic, directed towards the center of the pinning site, and given by
$
  \bm{F}_i^p = F_p \sum_j^{N_p} (\bm{r}_j^p - \bm{r}_i) \, \Theta(r_p - |\bm{r}_i - \bm{r}_j^p|).
$
For both $\bm{F}_i^{vv}$ and $\bm{F}_i^p$ the unit of length is $\lambda$.
The thermal forces $\bm{F}_i^T$ are modeled by Langevin kicks with the properties $\langle \bm{F}^T \rangle = 0$ and $\langle \bm{F}_i^T(t) \bm{F}_j^T(t') \rangle = 2 \eta k_B T \, \delta_{ij} \, \delta(t-t')$, where $k_B$ is the Boltzmann constant.
The Langevin kicks are implemented by applying random forces in the $x$ and $y$ directions drawn from a Gaussian distribution with zero mean and a standard deviation of $F^{T}$.
In our dimensionless units $F^{T}$ is the strength of the thermal forces and referred to as the temperature of the system.
The values of $r_p$ and $F_p$ permit at most one vortex to occupy each pinning site, and were chosen by trial and error in order to stabilize a metastable VL configuration yet still allow the system to be driven towards the equilibrium configuration.

In the two-dimensional MD approximation, the vortices are assumed to be stiff lines spaced far enough apart that their cores do not overlap.
This is appropriate as MgB$_2$ is a three dimensional (i.e., not layered) materials with weak pinning and therefore little or no vortex bending is expected.
It should also be noted that MD simulations have been used extensively to model the dynamics of stiff vortices in systems with pinning~\cite{Reichhardt.1995,Moon.1996,Fily.2010}.

\section{Results}
The results are organized as follows.
First, the protocol used to prepare the metastable VL configuration is described, as is how the bulk behavior is characterized by the peak splitting in the structure factor.
Second, it is demonstrated how the VL can be driven towards the equilibrium configuration by successive applications of the effective field oscillation, and the dynamics of this process is explored as a function of the oscillation amplitude.
Finally, the evolution of VL grain boundaries and energetics is studied at the local scale as the system is driven towards the equilibrium state.

We fix the twelvefold anisotropy amplitude at $K_{12} = 0.0125$ and vary the anisotropy ratio $\kappa$ by changing $K_6$, which was previously established to produce reliable results~\cite{Olszewski:2020jy}.
Varying $\kappa$ changes the magnitude of the interaction potential experienced by each vortex, which is equivalent to introducing a change in the effective vortex density.
To eliminate this variation, the effective magnetic field, defined by a two-dimensional integral of the interaction potential
\begin{equation}
  B_{\rm{eff}} \propto \int_0^{\infty} r \, dr \int_0^{2\pi} d\theta \; U(r,\theta) \propto A_v \left( 1 + \frac{K_6}{2} + \frac{K_{12}}{2} \right),
  \label{Beff}
\end{equation}
is kept at a constant value.
Here, $K_6 = 0$ and $A_v = 2.0$ is used as a reference, and the magnitude of the isotropic contribution to the interaction potential is adjusted such that $A_v = 2 (2 + K_{12})/(2 + K_6 +K_{12})$.

\subsection{Preparing a metastable VL}
\begin{figure}
	\includegraphics{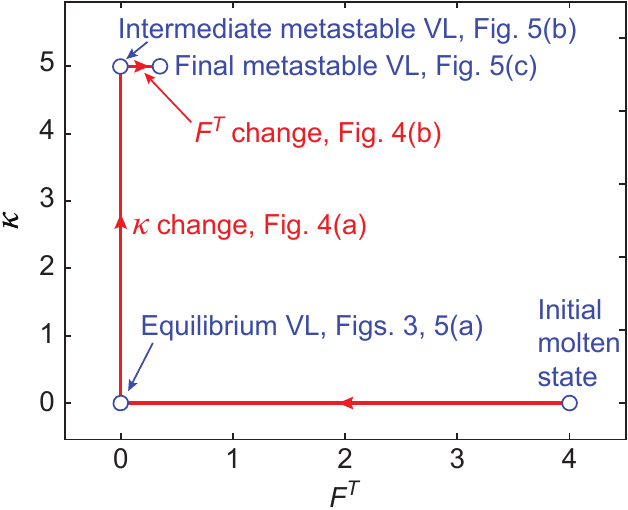}
    \caption{\label{fig2}
        Protocol used to prepare the metastable VL state, plotted as a function of anisotropy ratio $\kappa$ and temperature $F^T$.
        From a high temperature molten state at $\kappa = 0$, the temperature is decreased to zero, the anisotropy ratio is then increased to $\kappa = 5$, and finally the temperature is increased to $F^T = 0.35$.
        Labels indicate the locations and/or trajectories in parameter space illustrated in Figs.~\ref{fig3}, \ref{fig4} and \ref{fig5}.}
\end{figure}

The protocol used to prepare the VL in a metastable configuration is shown schematically in Fig.~\ref{fig2}.
The system begins with  $\kappa = 0$ and a temperature $F^T = 4.0$, which is sufficiently high to put it in a molten state.
    It is then cooled gradually to $F^T =0.0$ by reducing the temperature by $\Delta F^T = -0.05$ every $2 \times 10^4$ simulation time steps, which is sufficiently slow to ensure that the system reaches an equilibrium state~\cite{Olszewski:2020jy}.
The anisotropy ratio is then increased to $\kappa = 5$ in increments of $\Delta \kappa = 0.25$ every $2 \times 10^4$ simulation time steps, while keeping $F^T = 0$.
Finally, the temperature is increased to $F^T = 0.35$ using the same rate as for the initial cooling.
The final temperature increase provides the system with the necessary thermal energy to overcome activation barriers and be driven towards the equilibrium configuration by the effective field oscillations described in Sect.~\ref{SecEFO}.

\begin{figure}
    \includegraphics{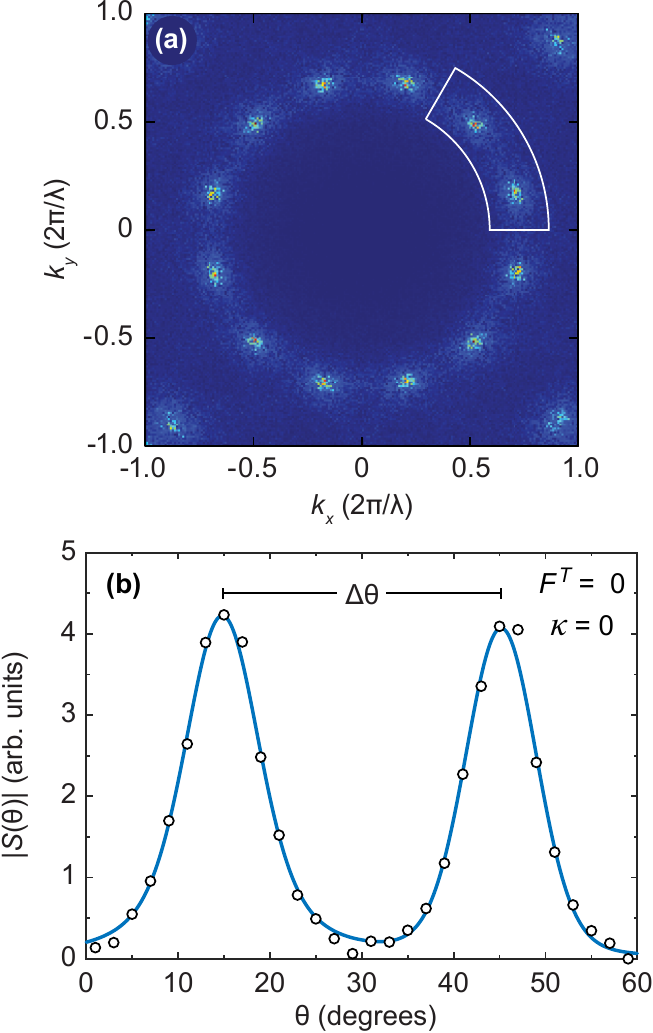}
    \caption{\label{fig3}
        Bulk characterization of the VL configuration.
        (a) False color plot of the structure factor amplitude $|S(\bm{k})|$ as a function of $k_y$ vs $k_x$ following an initial annealing to $F^T = 0$ with $\kappa = 0$.
        (b) Azimuthal dependence of the structure factor amplitude $|S(\theta)|$ folded into the $60^{\circ}$ segment indicated by a white outline in panel (a).
        The line is a fit to a two-peak Voigt function.}
\end{figure}

The global VL configuration is characterized by computing the structure factor, $S(\bm{k}) \propto |\sum^{N_v}_i \exp(-i \bm{k} \cdot \bm{r}_i)|^2$.
Figure~\ref{fig3}(a) shows $S({\bf k})$ at $\kappa = 0$ following the initial annealing to $F^T = 0$.
Here the intensity is concentrated in 12 (Bragg) peaks lying on a circle, indicating the presence of VL domains that are oriented along one of two different directions corresponding to the minima in the interaction potential in Eq.~(\ref{K612potential}).
The radius agrees with the calculated value for a triangular lattice, $k_0 = (2/\sqrt{3})^{1/2} \; 2\pi (\sqrt{5130}/108\lambda) = 0.71 (2\pi/\lambda)$.
Figure~\ref{fig3}(b) shows the azimuthal dependence of the structure factor folded into one $60^{\circ}$ segment, fitted by a two-peak Voigt function.
Voigt fits (a convolution of  a Gaussian function and a Lorentzian) are used throughout this work because they were found to provide better fits to the azimuthal intensity distribution than simple Gaussian or Lorentzian.
The peak splitting, $\Delta \theta$, is defined as the difference between the two fitted peak positions.

\begin{figure}
   \includegraphics{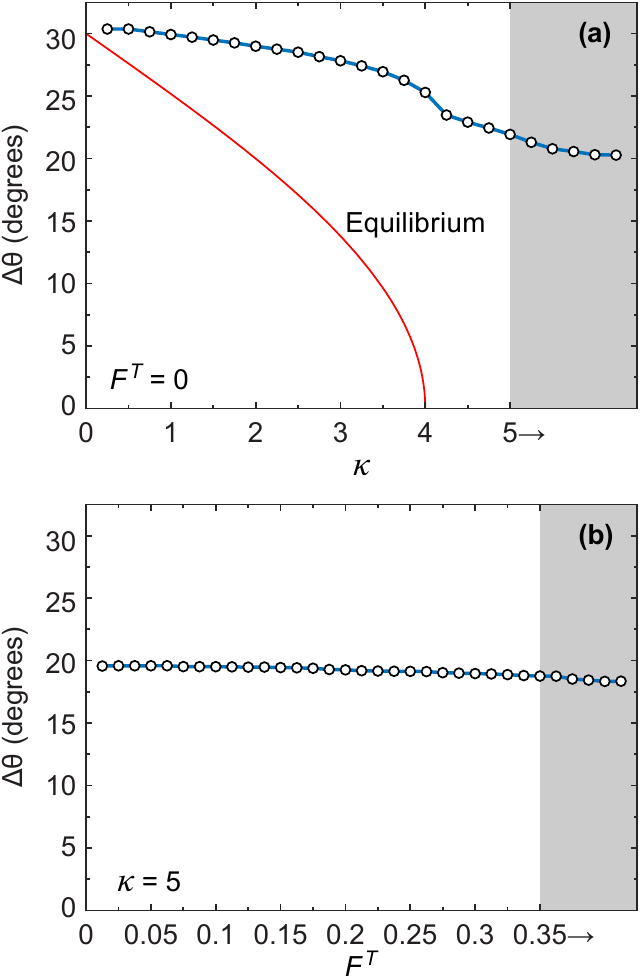}
    \caption{\label{fig4}
        Evolution of the structure factor peak splitting $\Delta \theta$ during the preparation of the metastable VL.
        (a) $\Delta \theta$ (circles) vs the anisotropy ratio as $\kappa$ is increased at zero temperature.
        After reaching $\kappa = 5$ the simulation was continued for an additional $2 \times 10^4$ simulation time steps (shaded area) with $\kappa$ fixed to allow the system to reach a static configuration.
        The red line indicates $\Delta\theta_{\text{eq}}$, expected for an equilibrium VL.
        (b) $\Delta \theta$ vs temperature under warming with fixed $\kappa = 5$.
        After reaching $F^T = 0.35$, the simulation was continued for an additional $2 \times 10^4$ simulation time steps (shaded area) with $F^T$ fixed to allow the system to reach a static configuration.}
\end{figure}

A metastable VL is established when $\kappa$ is increased at zero temperature, since without thermal fluctuations the system is no longer able to evolve enough to remain in the changing equilibrium state.
This is illustrated in the plot of $\Delta \theta$ in Fig.~\ref{fig4}(a).
Although $\Delta \theta$ decreases it always remains higher than the equilibrium value $\Delta\theta_{\text{eq}} = 60^{\circ} - 2 \theta_{\text{min}}(\kappa)$, indicating that the system becomes trapped in a metastable state.
The measured splitting angle has a kink near $\kappa = 4$ where $\Delta\theta_{\text{eq}}$ reaches zero.
Once the final value of $\kappa = 5$ has been reached, the system is allowed to evolve for an additional $2 \times 10^4$ simulation time steps.
This only gives rise to a modest additional reduction of the splitting angle, which stabilizes at a value of $\Delta \theta \sim 20^{\circ}$.
During the subsequent temperature increase at $\kappa = 5$, $\Delta \theta$ remains largely unchanged as shown in Fig.~\ref{fig4}(b), 
indicating that the thermal fluctuations alone are insufficient to permit the system to escape the metastable state.

\begin{figure*}
    \includegraphics{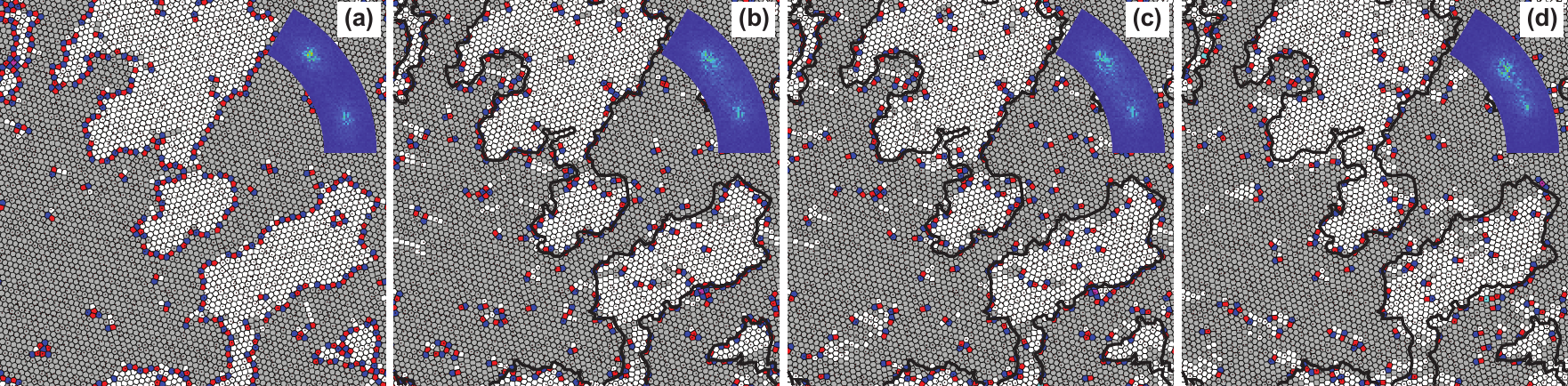}%
    \caption{\label{fig5}
        Voronoi constructions of the vortex positions for different system configurations.
        Red, white, and blue Voronoi cells correspond to five-, six-, and seven-fold coordinated vortices, respectively.
        The grey and white shading indicate the VL orientation.
        Inserts show the structure factor amplitude in the $60^{\circ}$ segment of reciprocal space indicated in Fig.~\ref{fig3}(a).
        (a) Equilibrium VL for $\kappa = 0$ and $F^T = 0$, following an annealing from the initial molten state.
        Domains are separated by well-defined grain boundaries decorated by 5-7 dislocations.
        (b) Metastable VL following an increase of the anisotropy ratio to $\kappa = 5$ at fixed $F^T = 0$, corresponding to the rightmost data point in Fig.~\ref{fig4}(a).
        Grain boundaries have fewer dislocations and some have shifted slightly.
        (c) Metastable VL following a subsequent heating to $F^T = 0$.35 with $\kappa = 5$, corresponding to the rightmost data point in Fig.~\ref{fig4}(b).
        Only minor changes are observed with respect to the zero temperature state in (b).
        (d) After the application of 1000 penetration depth oscillations at $F^T = 0$.35 and $\kappa = 5$.
        The number of dislocations is further reduced and more grain boundaries have shifted.
        Black lines in panels (b), (c) and (d) indicate the location of VL grain boundaries in (a).}
\end{figure*}

Real space information about the VL is obtained from Voronoi constructions of the vortex positions.
Within each Voronoi cell, all points are closer to a given vortex than to any other.
Furthermore, the number of edges yields the coordination of the vortex within the cell.
The Voronoi constructions provide local information about domain formation, VL defects and grain boundaries.
Figure~\ref{fig5}(a) shows the Voronoi construction for an equilibrium VL at $\kappa = 0$ and $F^T = 0$, corresponding to Fig.~\ref{fig3}.
Here the system breaks up into domains with two different orientations.
Within each triangular domain, the vortices are sixfold coordinated, but due to the twelvefold anisotropy of the vortex-vortex interaction potential, there are two degenerate orientations.
The grain boundaries between domains are decorated by dislocations consisting of fivefold and sevenfold coordinated vortices.

Figure~\ref{fig5}(b) shows the Voronoi construction plotted after the anisotropy ratio has been increased to $\kappa=5$ at zero temperature.
Although the number of dislocations has decreased, the system still shows clear domain formation with easily discernible grain boundaries.
Some grain boundaries have shifted relative to those in Fig.~\ref{fig5}(a) and indicated by the black lines.
Furthermore, the VL domains have rotated towards each other as indicated by a reduced peak splitting in the structure factor.
In other words, the relaxation towards the equilibrium configuration occurs primarily through a continuous rotation of the metastable VL domains rather than the nucleation and growth of separate ground state domains.
The ground state for this $\kappa$ would be a single domain containing sixfold coordinated vortices.
Heating the system to $F^T = 0.35$ at $\kappa = 5$ causes some additional annihilation of dislocations and grain boundary motion, but the VL remains in a metastable configuration as seen from the Voronoi construction and structure factor in Fig.~\ref{fig5}(c).
This confirms that thermal effects at this temperature are not strong enough to permit vortices to hop over the activation barrier between the metastable equilibrium states.

\subsection{Effective field oscillations}
\label{SecEFO}
In the SANS experiments, the transition from the metastable to the equilibrium state is induced by the application of magnetic field oscillations, which change the number of vortices in the sample~\cite{Louden:2019wx}.
Changing the number of vortices is not possible in the MD simulations with periodic boundary conditions, and instead an effective change in the vortex density is modeled by varying the penetration depth $\lambda = \lambda_0 + \Delta \lambda$.
This corresponds to a change in both the system size $L = 108 \lambda$ and the range of the vortex-vortex interaction in Eq.~(\ref{Fvv_a}) and (\ref{Fvv_b}) as $r$ is measured in units of $\lambda$.
As such, it is equivalent to a change of the magnetic field
\begin{equation}
    B = \frac{N_v \Phi_0}{(L \lambda)^2}
      = \frac{B_0}{(1 + \Delta \lambda/\lambda_0)^2},
\end{equation}
where $\Phi_0$ is the flux quantum and $B_0$ is the field for $\Delta \lambda = 0$.
The corresponding amplitude of the field variation, $\Delta B = B - B_0$, is given by
\begin{equation}
    \frac{\Delta B}{B_0} = \frac{1}{(1 + \Delta \lambda/\lambda_0)^2} - 1.
\end{equation}

Changing the range of the vortex-vortex interactions by varying $\lambda$ modifies the force balance between the unevenly spaced vortices near the grain boundaries.
This creates creates mechanical instabilities that allow for vortex jumps or rearrangements to occur.
It is important to note that whereas varying $\lambda$ in an experiment is associated with a change in temperature, all the effective field oscillations are performed with a constant $F^T$.
An alternative to the effective field oscillations would have been to drive the system would be with an applied current.
Such an approach was previously used to study the dynamical annealing of disordered vortex states~\cite{Daroca.2010}.

Each effective field oscillation is performed by changing $\lambda$ in steps of $0.05$ every 2,000 simulation time steps until the desired value of $\Delta \lambda$ is reached.
The cycle is completed by changing the penetration depth back to the original value $\lambda = \lambda_0$ at the same rate.
It was verified that this rate is slow enough that the results do not change if the frequency is further reduced.
Values of $\Delta \lambda$ between $-0.1 \lambda_0$ and $-0.5 \lambda_0$ were studied, corresponding to $\Delta B/B_0$ between $0.23$ and $3.0$.
Figure~\ref{fig5}(d) shows the Voronoi construction and structure factor after $1000$ penetration depth oscillations were applied to the mestastable state in Fig.~\ref{fig5}(c).
The number of dislocations are reduced and the VL domains have rotated closer to the equilibrium orientation as indicated by the smaller structure factor peak splitting.
Furthermore, more grain boundary motion is evident and also some fracturing of the VL into smaller domains.
This shows that the penetration depth oscillations are able to move the grain boundaries and thus mimic the behavior observed in the SANS experiments~\cite{Louden:2019wx}.
Notably, while all simulations were performed with a different configuration of the pin locations the system evolve toward the same equilibrium state.

One difference between the penetration depth oscillations, used in the MD simulations, and the experimental magnetic field oscillations is that the latter produces a small gradient in the vortex density perpendicular to the sample boundary.
This gradient gives rise to a current, which, in turn, causes vortex motion.
In comparison, no current is applied in the MD simulations since the simulations use periodic boundary conditions and there is no sample boundary.
It is therefore not possible to directly compare the values of the effective field oscillations used in the MD simulations and the much smaller values used in the SANS experiments ($\Delta B/B_0 = 0.001 - 0.003$)~\cite{Louden:2019wx}.

\subsection{Metastable state relaxation dynamics}
\begin{figure}
    \includegraphics{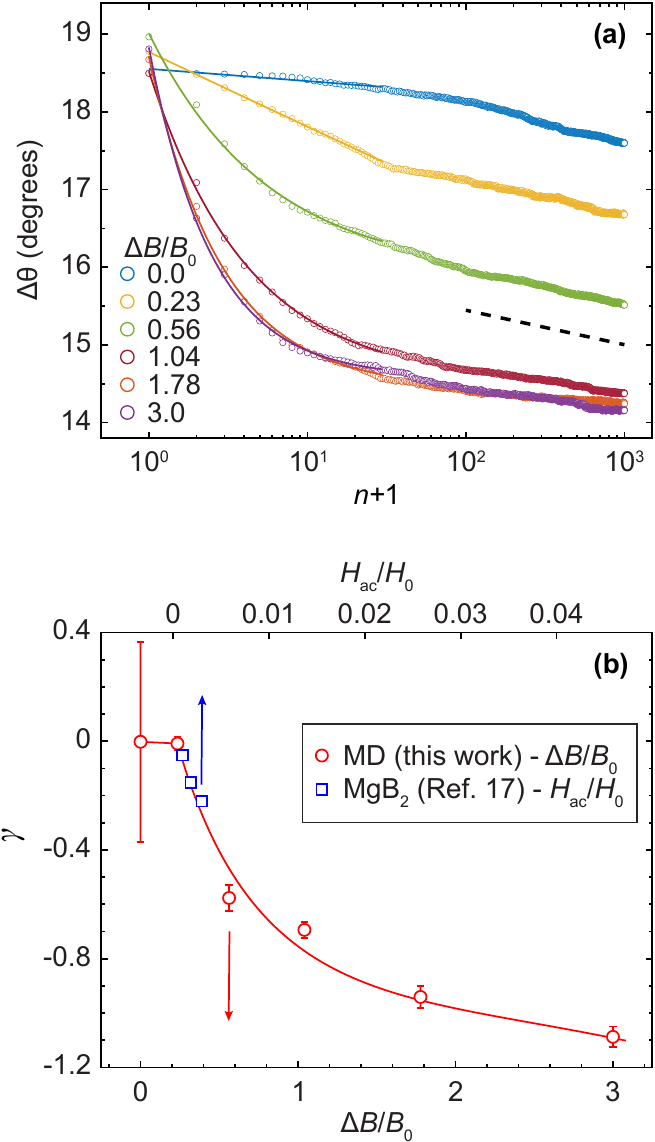}
    \caption{\label{fig6}
        Metastable state relaxation dynamics.
        (a) Structure factor peak splitting $\Delta \theta$ versus the penetration depth cycle count $n+1$.
        The data are an average over ten simulation runs.
        The solid lines are fits to the form $\Delta \theta = a(n+1)^\gamma + b$ for $n+1 \leq 25$ .
        For large $n$, all data show a logarithmic behavior as indicated by the dashed line.
        (b) The exponent $\gamma$ from the fits in (a) versus $\Delta B/B_0$.
        Also shown for comparison is the exponent versus ac field amplitude $H_{\text{ac}}/H_0$, obtained from SANS experiments on MgB$_2$~\cite{Louden:2019wx}.
        The line is a guide to the eye.}
\end{figure}

To study the relaxation dynamics of the metastable VL state, MD simulations were performed in which the system was subjected to successive penetration depth oscillations.
Simulations were performed for $\Delta B/B_0 = 0, 0.2346, 0.5625, 1.0408, 1.7778,$ and $3.0$.
Ten simulation runs were performed for each value of $\Delta B/B_0$.
Figure~\ref{fig6}(a) shows $\Delta \theta$ determined from the structure factor peak splitting, averaged over the ten simulation runs, versus the cycle count $n+1$, where $n$ is the number of penetration depth oscillations that have been applied.
Using $n+1$ as the abscissa allows the inclusion of the initial metastable state ($n = 0$).

For non-zero $\Delta B/B_0$ an initial rapid drop of $\Delta \theta$ is observed, with a magnitude that increases with the amplitude of the effective field oscillation.
The data for $n \leq 25$ is well fitted by the functional form $\Delta \theta = a(n+1)^\gamma + b$, as shown by the full lines in Fig.~\ref{fig6}(a).
A similar form was also used to fit the experimental SANS data~\cite{Louden:2019wx}.
Fitted values of $\gamma $ as a function of $\Delta B/B_0$ are shown in Fig.~\ref{fig6}(b).
For $\Delta B/B_0 = 0$ (no oscillations), $\Delta \theta$ decreases very slowly due only to thermal effects at $F^T = 0.35$ and $\gamma\sim 0$ although the uncertainty is large.
For the lowest non-zero $\Delta B/B_0$, $\gamma$ is still very small but the uncertainty is significantly reduced.
As the amplitude of the field oscillations is increased, the magnitude of $\gamma$ also grows until, for the highest value of $\Delta B/B_0$, $\gamma = -0.5$.
The overall behavior of $\gamma$ with increasing $\Delta B/B_0$ is similar to that found in the SANS experiments~\cite{Louden:2019wx}, 
where exponents ranging from $\gamma=-0.051$ to $\gamma=-0.22$ were obtained as shown in Fig.~\ref{fig6}(b).
It is likely that the largest values of $\Delta B/B_0$ used in the MD simulations exceed the range explored in the SANS in the experiments, but as discussed a direct comparison is not possible.
As evident from Fig.~\ref{fig6}(a), the power law fits of $\Delta \theta$ describe the data well for $n \lesssim 300$; however, at large $n$ they do not perform as well.
Rather, the system exhibits a logarithmic behavior at large $n$ for all values of $\Delta B/B_0$.

\begin{figure}
    \includegraphics{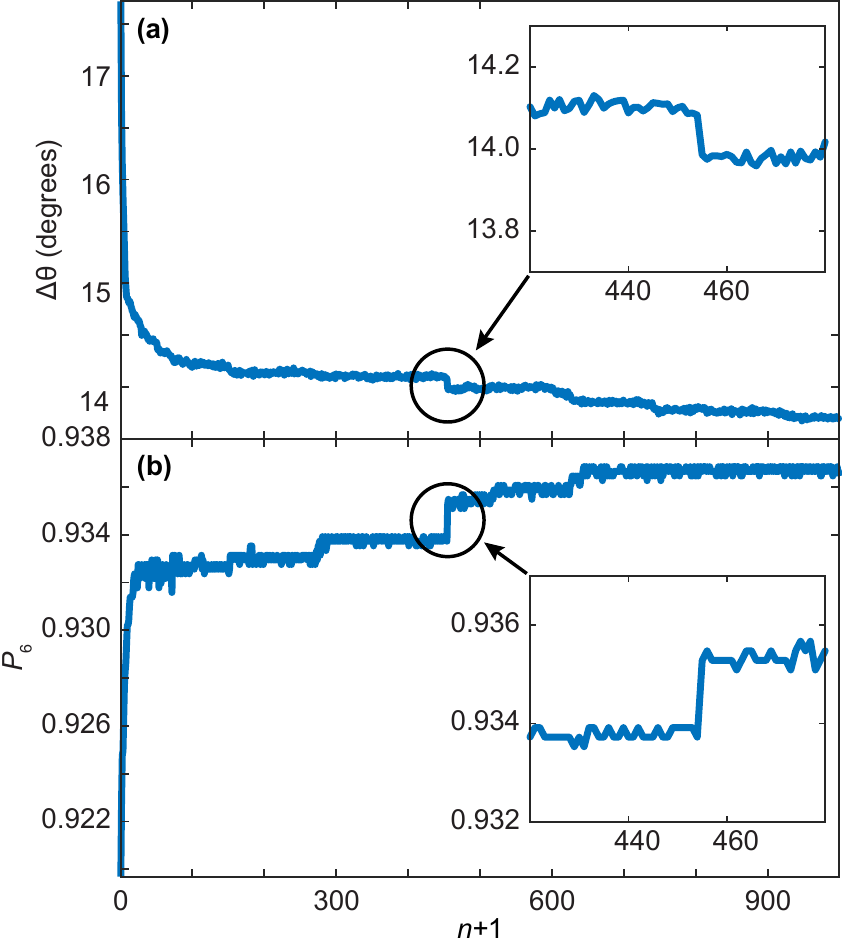}
    \caption{\label{fig7}
        A single simulation run showing sudden, stepwise relaxations for $n \gtrsim 200$.
        (a) Splitting angle $\Delta \theta$ versus $n+1$ for $\Delta B/B_0 = 3.0$.
        This shows a series of discontinuous jumps down at large $n$. 
        (b) Corresponding fraction of sixfold coordinated particles $P_6$.
        The sudden decreases in the splitting angle are accompanied by increases in $P_6$, corresponding to the annihilation of dislocations. 
        The insets highlight the jump near $n+1 = 450$.}
\end{figure}

To gain a better understanding of the longer time behaviors, Fig.~\ref{fig7}(a) shows $\Delta \theta$ for $\Delta B/B_0 = 3.0$ versus $n+1$ on a linear scale for a single simulation run. 
After the initial smooth decrease, a series of sharp downward jumps or avalanches appears for $n > 100$.
Figure~\ref{fig7}(b) shows the corresponding fraction of sixfold coordinated vortices $P_6$ versus $n+1$.
From this it is evident that the downward jumps in $\Delta \theta$ are accompanied by sudden increases of $P_6$, meaning that the avalanches are due to the annihilation of groups of dislocations.
This is different from the behavior for $n+1 \leq 15$ where $P_6$ undergoes a smooth continuous increase.

\begin{figure}
    \includegraphics{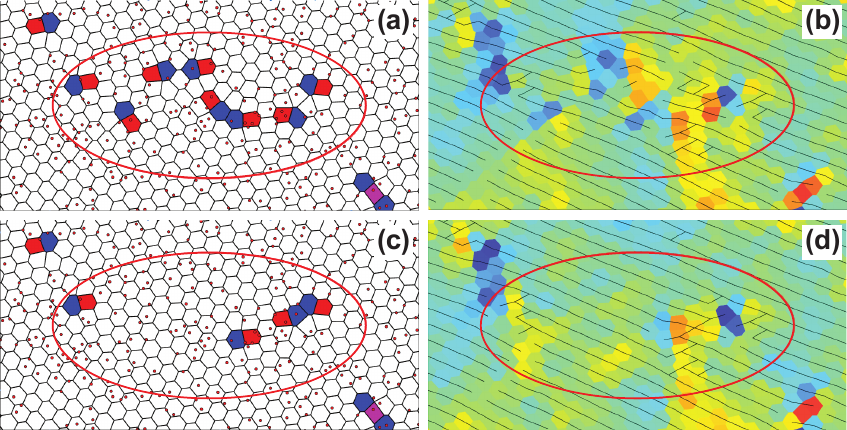}
    \caption{\label{fig8}
        Portion of the sample (a,b) just before and (c,d) after the jump in $\Delta \theta$ and $P_6$ near $n+1 = 450$ in Fig.~\ref{fig7}.
        The circled region indicates the location of a VL grain boundary.
        (a,c) Voronoi constructions of the vortex positions.
        Colors are the same as in Fig.~\ref{fig5}.
        (b,d) Corresponding heatmaps showing the local energy of each vortex based on its interactions with the surrounding VL.
        Yellow and red is higher than the average and blue is lower.
        The lines give an approximate indication of the local domain orientation.}
\end{figure}

A more direct demonstration that the jumps in the splitting angle occur by means of avalanches, in which multiple dislocation pairs are annihilated, is shown in Fig.~\ref{fig8}.
This shows a portion of the sample immediately before and after the jump highlighted in the insets of Fig.~\ref{fig7}.
In the Voronoi construction prior to the jump in Fig.~\ref{fig8}(a), eight 5-7 dislocations are located near a VL grain boundary within the circled region.
After the jump, in Fig.~\ref{fig8}(c), about four of the dislocations 
have annihilated.
This is also reflected in the local VL energy shown in Figs.~\ref{fig8}(b) and \ref{fig8}(d).
After the jump, the overall energy of the system has dropped and become more uniform.

In summary, the dynamics can be broken into two parts with a fast initial power law dependence on $n$ followed by a slower universal logarithmic dependence dominated by bursts or avalanches.

\section{Discussion}
The decay of a metastable state studied here and in experiments could have implications for many other non-thermal systems in metastable states.
One example is granular matter confined in a vertical container.
After the grains are initially poured in to the container, they have a fixed volume fraction; however, if the container is subjected to periodic vertical shaking or tapping, the grains become more compressed and the volume fraction $V$ varies as $V \propto 1/(a + b \log(n))$ where $n$ is the number of taps applied~\cite{Nowak98}.
In 2D systems, there have been several studies examining the coarsening dynamics of topological defects in the case where the equilibrium system would be defect free.
Over time, the density of defects decays according to a power law with an exponent in the range $1/4$ to $1/3$~\cite{Elder92,Purvis01,Boyer02,Boyer04}.
The values of the exponents are consistent with some of the values we obtain at shorter times. 
Studies of systems that form 2D crystals in the absence of quenched disorder show that when quenched disorder is added, numerous defects appear in the lattice that form grain boundaries at zero temperature; however, if an ac driving is applied that shakes the particles back and forth over the substrate, the defects 
annihilate over time at a rate that becomes proportional to $\ln(t)$ at longer times~\cite{Valenzuela02,Wei04,Reichhardt06a}.
If quenched disorder is present, there are generally some defects that remain permanently trapped.

One of the closest systems outside of superconducting vortices to which we can compare our results is the mechanical annealing of grain boundaries in 2D colloidal crystals, where experiments found that the density of grain boundaries decays as a power law as a function of time~\cite{Wei04}.
It may be possible that the coarsening dynamics has a faster time scale than the glassy dynamics which are typically associated with a $\ln(t)$ dependence, so that for the first 1 to 100 cycles the system behaves as if it is coarsening, but at longer times it reverts to $\ln(t)$ behavior when the dynamics becomes dominated by hopping or avalanches.

Although our results are not accurate enough to distinguish between the different types of coarsening, at early times they do show that the basic microscopic ingredients for the model we have presented capture the main features found in the experiments.
Additionally, our results agree with the vortex-vortex interactions observed in the experiments, where a metastable twelvefold symmetric state is quenched into a sixfold symmetric state and the ac driving progressively reduces the splitting angle.

It could also be possible that in some cases an ordered system in the presence of quenched disorder would show increased defect formation under an applied drive because
some particles become trapped while other particles shear past them.
This also suggests that there could be an optimal ac amplitude or frequency for creating the maximum amount of dynamic annealing.
In addition, introduction of quenched disorder can cause the system to remember the metastable states.
If a cyclic driving current were applied, the vortices would move in a periodic manner, and it would be possible to examine whether they return to the same positions at the end of each cycle in a form of return point memory or random organization, which has been studied in numerous periodically driven systems~\cite{Pierce05,Corte08,Libal12,Pasquini21}.

\section{Summary}
The relaxation of a metastable vortex state was studied numerically for a system quenched from a twelvefold to a sixfold anisotropy of the vortex-vortex interaction.
The system is initially in an equilibrium configuration for the twelvefold anisotropy, consisting of triangular VL domains with two different orientations and separated by grain boundaries decorated by dislocations.
In comparison, the ground state for the sixfold anisotropy is a single trianglar VL domain.
The bulk VL configuration is characterized by the structure factor.
Following the quench from the twelvefold to the sixfold anisotropy, the initial twelve equally spaced structure factor peaks form six pairs with a decreasing splitting.
However, the presence of grain boundaries and weak pinning traps the system in a metastable state with a non-zero splitting angle.

The system is driven towards the ground state by a periodic variation of the penetration depth, which mimics the magnetic field oscillations used in SANS experiments~\cite{Louden:2019wx}.
This causes a gradual decrease of the structure factor peak splitting with successive applications of the penetration depth cycles.
Initially, the decrease follows a power law with an exponent that increases with the amplitude of the penetration depth oscillations, in agreement with the SANS experiments performed in a similar regime~\cite{Louden:2019wx}.
As the number of penetration oscillations grows, the decay of the peak splitting becomes logarithmic with a slope that is independent of the amplitude.
This indicates a transition from annihilation of individual dislocation pairs at short times to coarsening dynamics at longer times, where the splitting angle decreases in bursts or avalanches that are correlated with the annihilation of larger groups of dislocations.

\section*{Acknowledgements}
We are grateful to E.~Roe, M.~W.~Olszewski and D.~Spulber for assistance with the molecular dynamics simulations and analysis of the results.
This research was supported in part by the Notre Dame Center for Research Computing.
Work at the University of Notre Dame (DM, MRE: MD simulations, data analysis) was supported by the US Department of Energy, Office of Basic Energy Sciences, under Award No. DE-SC0005051.
Part of this work (CR, CJOR: code development) was carried out under support by the US Department of Energy through the Los Alamos National Laboratory.  Los Alamos National Laboratory is operated by Triad National Security, LLC, for the National Nuclear Security Administration of the U. S. Department of Energy (Contract No. 892333218NCA000001).

\section*{References}
\bibliography{bib}

\end{document}